\documentclass[aps,prl,groupedaddress,showpacs,twocolumn,floatfix]{revtex4}
\usepackage{amsmath}
\usepackage{amssymb}
\usepackage{graphicx}
\usepackage{epsfig}

\begin{document}

\title{Vortices in a rotating BEC under extreme elongation}
\author{P. S\'anchez-Lotero and J. J. Palacios}
\affiliation{\mbox{Departamento de F\'{\i}sica Aplicada, Universidad de Alicante\\ San Vicente del Raspeig, Alicante 03690, Spain}
}
 
\begin{abstract}
We investigate a non-axisymmetric rotating BEC in the limit of rotation frequency for which the BEC transforms into a quasi-one-dimensional system. We compute the vortex lattice wavefunction by minimizing the Gross-Pitaevskii energy functional in the lowest Landau level approximation for different confinement potentials. The condensate typically presents a changing number of vortex rows as a function of the interaction strength or rotation-confinement ratio. More specifically, the vortex lattices can be classified into two classes according to their symmetry with respect to the longitudinal axis. These two classes correspond to different local minima of the energy functional and evolve independently as a function of the various parameters. 
\end{abstract}

\maketitle
\narrowtext

\section{Introduction}
In recent years several groups have studied theoretically and experimentally vortices in rotating Bose-Einstein condensates (BEC's). Butts and Rokhsar\cite{Butts} were the first in, with the help of the Gross-Pitaevskii (GP) functional, calculating the wave function of a rotating Bose-Einstein condensate using a lowest Landau level (LLL) approximation\cite{Landau}. Interestingly, a very similar approach was being developed at the same time to study a class of closely related systems: Vortices in superconducting disks\cite{Palacios}. The theoretical idea for inducing the mechanical rotation and coupling of internal states using an electromagnetic field (phase imprinting) in the condensate was proposed later by Williams and Holland\cite{Williams} and the experimental confirmation was given shortly afterwards by the JILA group\cite{Matthews}. Arrays of few singly quantized vortices were produced later with optical spoon stirring method\cite{Madison}. After these works, it was established that in a rotating BEC the energetically favorable state was a triangular lattice of vortices similar to the Abrikosov lattice in superconductors\cite{Abo}. Ho, based on this, showed that a three dimensional BEC with a big number of vortices can be described like a quantum-Hall system in the LLL\cite{Ho}. However, in the ''ultrafast'' rotating limit, where the rotation frequency $\Omega$ is close to the trapping frequency $\omega$, the GP mean field approximation fails\cite{Cooper}. The study of vortices in the ultrafast limit can still be done including a quartic radial potential in the GP functional which further confines the bosons and allows a mean-field description of the BEC above the critical rotation frequency\cite{Fetter}. This kind of trapping potential has allowed to find experimental evidences of the melting of the vortex lattice\cite{Bretin}, predicted in previous works\cite{Fetter,Cooper}.

\section{The Gross-Pitaevskii functional in the Landau gauge}
The vortex lattice state of a BEC in symmetric potential traps has been well studied with Gross-Pitaevskii and Thomas-Fermi formalisms\cite{monton}. Recently, however, the behavior of the vortex lattice in an anisotropic potential trap has caught the attention of investigators\cite{Engels} and interest to look for unsual vortex structures in BEC's under different conditions has risen\cite{Mueller}. Anisotropic BEC's are modeled adding an asymmetric potential to the usual harmonic confining potential which elongates the BEC. To find the wave function $\Psi_0$ of a rotating and quasi-two-dimensional BEC\cite{Haljan} confined by a symmetric harmonic potential one minimizes the GP functional\cite{Gross}
\begin{equation}
E[\Psi_0]=\int dV \left(E_k+E_h+E_i+E_r \right)
\end{equation} 
where $E_k=\frac{\hbar^2}{2m}|\vec{\nabla}\Psi_0|^2$ is the kinetic term, $E_h=\frac{m \omega_0^2 (x^2+y^2)}{2}|\Psi_0|^2$ is the harmonic term, $E_i=\frac{g}{2}|\Psi_0|^4$ is the interaction term, and $E_r=- \Psi_0^* \vec{\Omega} \cdot \vec{L} \Psi_0$ is the rotating term, with $\vec{\Omega}$ the rotating angular frequency, and $\vec{L}$ the angular momentum operator. We assume throughout an oblate BEC of witdh $a_z$ in the $z$ direction which keeps $\Psi_0$ in the ground state for this direction. 
We take $\vec{\Omega}=\Omega \hat{z}$ so the rotating term becomes 
\begin{equation}
-\Omega\Psi_0^* L_z \Psi_0= i\hbar\Omega\Psi_0^*\left[x\frac{\partial}{\partial y}-y\frac{\partial}{\partial x}\right]\Psi_0.
\end{equation}
We now consider the anisotropic trap which produces the elongation of the BEC. This trap is modeled by adding
\begin{equation}
E_{an}[\Psi_0]=\int dV \frac{m \Lambda^2 (x^2-y^2)}{2}|\Psi_0|^2,
\end{equation}
with $\Lambda$ varying between 0 and $\omega_0$ \cite{Rosenbusch,Linn}.
The total trapping potential is thus $V_{tot}=\frac{m}{2} [x^2(\omega_0^2+\Lambda^2)+ y^2(\omega_0^2-\Lambda^2)]$. When $\Lambda$ grows, the system becomes more confined in the $x$ direction and more elongated in the $y$ axis. These kind of systems have been studied theoretically in previous works for slightly elongated rotating traps\cite{Linn,Oktel} and, recently, for extreme elongation as well\cite{Sinha}. Setting $\Lambda^2= \omega_0^2 - \Omega^2$, the system losses the $y$ confinement and the BEC resembles a channel of length $l$, which tends to infinity, with parabolic confinement in the $x$ direction. Using the gauge transformation $\Psi_0=\Psi e^{-\frac{i}{\hbar}m \Omega xy}$\cite{Sinha} the energy functional can be written as
\begin{equation}
E[\Psi] = \int dV \Psi^* \left( -\frac{\hbar^2}{2m}\nabla^2 + V_{eff} + \frac{g}{2} |\Psi|^2 + 2i\hbar\Omega x\frac{\partial}{\partial y} \right) \Psi.
\end{equation}
where $V_{eff}=mx^2(\Omega^2 + \omega_0^2)$ is the effective trapping potential.
The relation between $\Lambda$ and $\Omega$ defines the existence of only one rotating critical frequency for a given anisotropy for which the BEC is just deconfined in $y$.

We use the oscillator length $a_0=\sqrt{\frac{\hbar}{m\omega_0}}$ and the harmonic ground state energy $E_0=\frac{\hbar\omega_0}{2}$ as length and energy units, respectively. We propose an expansion of the wavefunction $\Psi$ as a linear combination of lowest level eigenfunctions, $\phi_k$, of the kinetic energy operator, the rotating term, and the trapping potential:
\begin{equation}
\Psi=\sum_{k} C_k \phi_k.
\end{equation}
By symmetry we propose eigenfunctions of the form
\begin{equation}
\phi_k=Ae^{iky}\chi_k(x),
\end{equation}
where $A$ is a normalization constant. The one-dimensional differential equation for $\chi_k$  becomes
\begin{equation}
-\frac{d^2}{dx^2}\chi_k + \left[ \left(\sqrt{2 R}x-\frac{2}{\sqrt{2R}}\frac{\Omega}{\omega_0}k \right)^2 - \epsilon^*_k \right] \chi_k = 0,
\end{equation}
with  $\epsilon^*_k=\epsilon_k-k^2\frac{\left( 1-\frac{\Omega^2}{\omega_0^2} \right)}{\left( 1+\frac{\Omega^2}{\omega_0^2} \right)}$, $\epsilon_k$ being the eigenenergies of the eigenfunctions $\phi_k$, and $R=1+\frac{\Omega^2}{\omega^2_0}$.
We find the set of normalized functions $\chi_k$ numerically and analytically and set the normalization constant $A$ to $\frac{1}{\sqrt{V}}$ with $V=a_0la_z$. In the rotating frame the functions $\phi_k$ are similar to the solutions of the Schr\"{o}dinger equation for a charged particle in presence of a magnetic field described by a vector potential in the Landau gauge\cite{Chakaborty}. An analysis of the analytical solution of the differential equation can be found in the Appendix I.

Now the GP functional becomes
\begin{eqnarray}
& &E=\sum_{k}^{N_c} |C_{k}|^2 \epsilon_k + \frac{g}{2} \sum_{k_1,k_2,k_3,k_4}^{N_c} C_{k_1}^{*}C_{k_2}^{*}C_{k_3}C_{k_4} \nonumber \\
& &\int dx \chi_{k_1} \chi_{k_2} \chi_{k_3} \chi_{k_4} \delta_{k_3+k_4,k_1+k_2}
\end{eqnarray}
with $E$ in units of $2\pi \frac{a_0}{l}\frac{\hbar \omega_0}{2}$, $g$ in units of $\frac{\hbar \omega_0}{2}V$, and $N_c$ being the number of terms in the linear combination. The first part of the GP functional includes the kinetic, harmonic and rotating terms of the functional.
The above expansion turns out to be  similar to the one used by Abrikosov in his description of the vortex lattice superconductors\cite{Abrikosov}. The Abrikosov lattice in superconducting strips was described using a similar expansion by one of the authors\cite{Palacios}. Here we make use of our previous experience to minimize semianalytically the GP functional with respect to the coefficients and the wave vectors $k$, imposing the conservation of particles constraint
\begin{equation}
\int dV \left( |\Psi|^2 - n \right)=0,
\end{equation}
where $n$ is the particle density. For an extended analysis of the minimization details see Appendix II.

\begin{figure}
\centering
\includegraphics[scale=0.7]{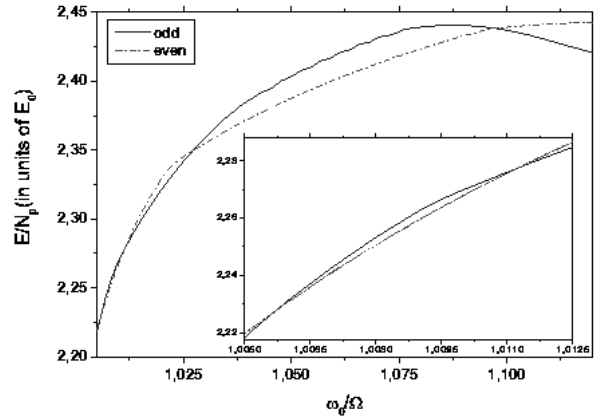}
\caption{Energy as a function of confinement for the two families of configurations: even (dashed-dotted) and odd (solid). The ground state corresponds to the lower envelope.\label{fig3}}
\end{figure}

\section{Vortex lattices in extremely elongated BEC}
We begin by addressing the case of purely parabolic confinement. An interaction of $gN=1.91$ is fixed throughout. We minimize analytically when feasible (for up to $N_c=3$), while for a higher number of terms we resort to numerical minimization. Next we add an infinite well to the parabolic confinement to simulate a deviation from non-harmonicity. If the infinite well is thinner than the lateral extension of the BEC, the bosons feel the additional confinement. The existence of the infinite potential allows us to study the BEC for rotation frequencies higher than the deconfinement frequency in the spirit of Ref.\cite{Fetter} for axisymmetric potentials. According to our results presented below, the behavior between elongated BEC's trapped by only harmonic potentials and BEC's confined by stronger confinement potentials is noticeably different even for rotation frequencies smaller than the critical one.

\subsection{Parabolic confinement}
\begin{figure}
\centering
\includegraphics[scale=0.2]{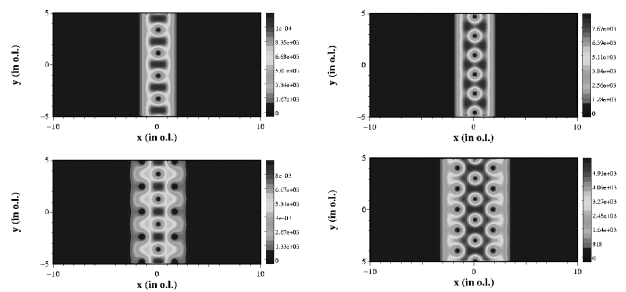}
\caption{Four examples of configurations found for the even family at different values of $\frac{\omega_0}{\Omega}$: 1.075, 1.029, 1.021 and 1.008. Lengths are in oscillator units (o.l.) and $gN=1.91$. The state in the third panel does not correspond to a ground state.
\label{even}}
\end{figure}

\begin{figure}
\centering
\includegraphics[scale=0.2]{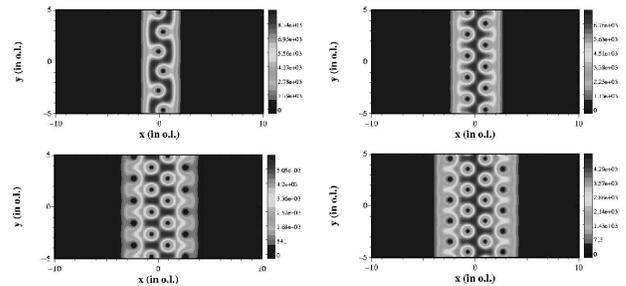}
\caption{Same as in Fig. 2 for the odd family at different values of $\frac{\omega_0}{\Omega}$: 1.075, 1.021, 1.008 and 1.006. The first and the third states do not correspond to ground states.
\label{odd}}
\end{figure}

In this section we show the behavior of an extremely elongated BEC as a function of the ratio $\frac{\omega_0}{\Omega}$ or the anisotropy $\Lambda$ (notice that they are related to maintain the condition for critical deconfinement in the  $y$ direction).  In general terms (which we detail below) we find two different families or classes of configurations (solid and dashed-dotted lines in Fig.\ \ref{fig3}). Those denominated "even" from now on have a vortex row in the center of the BEC and present even parity with respect to an imaginary line along the main BEC axis. In other words, they exhibit mirror symmetry, being invariant under the transformation than interchanges right and left. Those without a vortex row in the middle (denominated "odd" from now on) do not present mirror symmetry, but are invariant under a combined left-right interchange and a translation by the intervortex distance. Both classes of solutions are local minima of the multidimensional energetic landscape and evolve independently as the confinement varies. In other words, their different symmetry acts as a topological characteristic that keeps the two minima apart from each other in the energetic landscape, alternating in the ground state. This topological characteristic behaves in this sense as the vorticity in axisymmetric BEC's\cite{Butts} or superconducting disks\cite{Palacios}. Within each class of solutions, vortices enter the BEC progressive and continuously from the edges as the rotation increases. In general terms one can see that solutions in which the vortices are well-defined can be ground states while those where incipient vortices appear on the edge of the BEC are typically not ground states (see Figs. \ref{even} and \ref{odd}). The energy and its derivative are always continuous for the odd family with some exceptions for the even family that we describe below. Derivative discontinuities for this family sometimes occur in the ground state and add up to the ones that occur when the energy of the different classes cross each other (see Fig. \ref{fig3}). At the crossing points the ground state changes its topological character by increasing the number of vortex rows by one.

\begin{figure}
\centering
\includegraphics[scale=0.7]{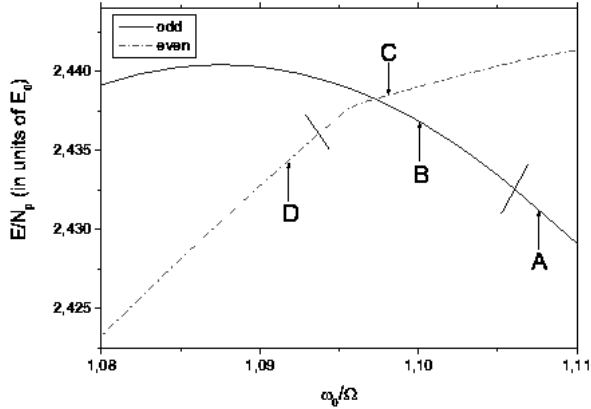}
\caption{Transition from no rows (solid line) to one vortex row (dashed-dotted line). Interaction strength is $gN=1.91$. Additional lines separate different configurations within the same family. The corresponding densities are shown in Fig.\ \ref{dens1}}
\label{fig1}
\end{figure}

\begin{figure}
\includegraphics[scale=0.15]{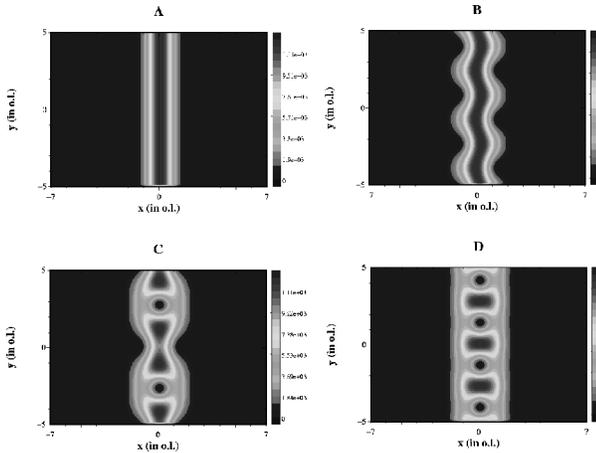}
\caption{Panels A, B, C, and D correspond to BEC densities at the points shown in \ref{fig1}. Lengths are in oscillator units (o.l.). The transition from zero (A, $N_c=1$) to one vortex row (D, $N_c=2$) presents two intermediate type of states (B, $N_c=3$ and C, $N_c=4$).}
\label{dens1}
\end{figure}

In Fig. \ref{fig1} we present a zoom of the first crossing point. For low rotation frequencies, the BEC ground state is just formed by the $k=0$ component in the expansion and does not contain vortex rows (A in Fig. \ref{dens1}). An excited state with a very sparse vortex line can also be found at those frequencies (C). In the evolution from a BEC with no vortex rows to a BEC with one vortex row, the ground state first develops a corrugation (B). This corrugation is the precursor of the nucleation at the edge of two vortex rows (see Fig.\ \ref{odd}). The states with two vortex rows are not the ground state until the vortices are fully developed (see below) at higher rotation frequencies. The appearance of the corrugated state presents a continuous derivative (signaled by the long bar in \ref{fig1}), i.e., the system presents a second-order phase transition before the row of vortices suddenly penetrate when the energy of the even family becomes lower and the solution with one vortex row becomes the ground state. The second-order phase transtion is not surprising since the symmetry that characterizes the odd class does not change as the corrugation developes. The corrugated states shown in Fig.\ \ref{dens1} can be formed by up to five terms. The appropiate variational flexibility is here crucial for the correct description of this transition. 

As mentioned, derivative discontinuties not only occur at the crossing points where the ground state changes its number of rows and its symmetry. Once the vortex row has appeared in the ground state, the sparse linear density of vortices becomes suddenly more dense (D) in a first-order phase transition (short bar in Fig.\ \ref{fig1}).  As far as we have been able to check in our numerics, this unexpected sudden change in the ground state is not an numerical artifact, although we cannot discard that a more complete expansion of the GP wave-function could smooth this transition into a second-order one. 

\begin{figure}
\includegraphics[scale=0.7]{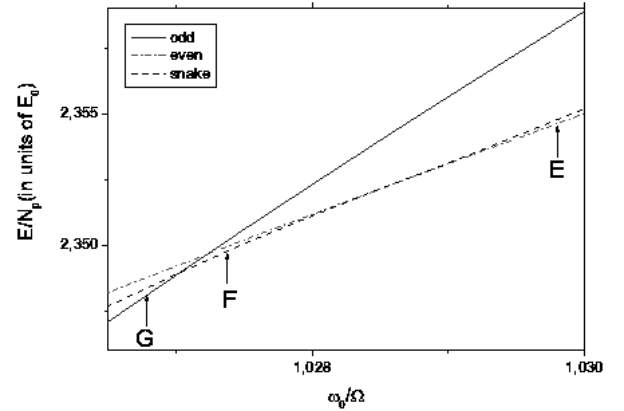}
\caption{Same as in Fig.\ \ref{fig1}, but for the transition from one row (dashed-dotted line) to two vortex rows (solid line). Notice the crossing intermediate state (dashed line, $N_c=5$).}
\label{fig2}
\end{figure}

\begin{figure}
\includegraphics[scale=0.11]{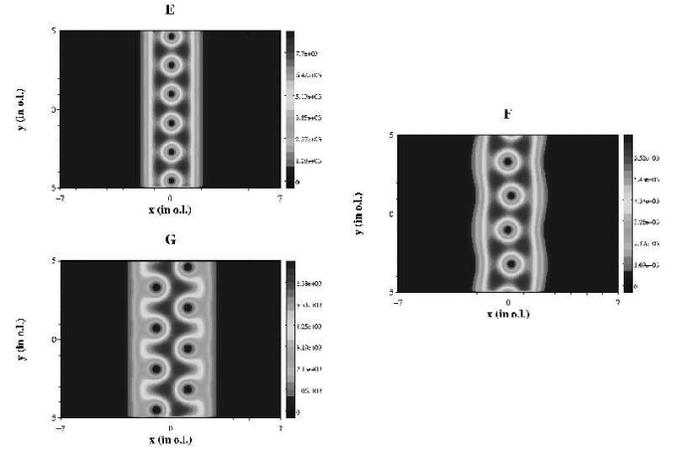}
\caption{BEC densities for the points E, F, and G in Fig.\ \ref{fig2}. Lengths are in oscillator units (o.l.).}
\label{dens2}
\end{figure}

The entrance of a second vortex row at the second crossing point also presents some peculiarities. As seen in Fig. \ref{fig2}, before the crossing with the odd family occurs again, a solution with one vortex row, but different symmetry from the ground state (E), becomes the ground state (F). This new solution has a snake-like form. The fact that the symmetry is different explains why the transition is of first order. This solution is, however, almost inmediately overcome by the two-row solution (G). From there on the number of rows increases one by one as shown in Figs. \ref{fig3}, \ref{even}, and \ref{odd}. Numerical limitations prevent us from making a thourough study of the crossing points as we have done in the first two, but we expect similar intermediate solutions to appear near the crossing points. 

\subsection{Parabolic + Infinite well confinement}
\begin{figure}
\centering
\includegraphics[scale=0.7]{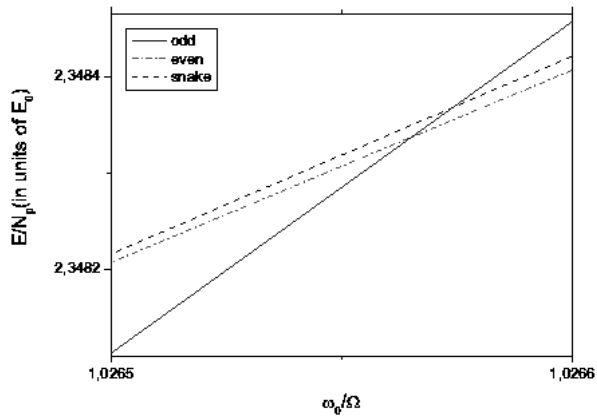}
\caption{Transition from one to two vortex rows for a quasi-parabolic potential defined by a parabolic and an infinite well of  width $7a_0$. The zig-zagging configuration is not a ground state for any rotation frequency. The correspondent densities are similar to the densities in Fig. \ref{dens2}. \label{fig4}}
\end{figure}

Fixing the interaction energy $gN=1.91$ and including an infinite well of width $7a_0$ we find a similar zero-one vortex row transition behavior for a non-parabolic confinement. This is expected since the dispersion relations for the parabolic and quasi-parabolic cases are fairly similar in the center of the condensate (for the chosen width). For a small rotation frequency the vectors $k$ of the expansion place themselves near the center and are nearly the same as those for the parabolic case. However for narrower wells, i. e., $4a_0$, the dispersion relation deviates strongly from the parabolic case due to the presence of the infinite potential. For the $7a_0$ system we find that the one-two vortex row  transition does not show the snake-like intermediate ground state. This solution has always a greater energy than the ground states with two vortex rows as shown in Fig. \ref{fig4}.

\begin{figure}
\centering
\includegraphics[scale=0.7]{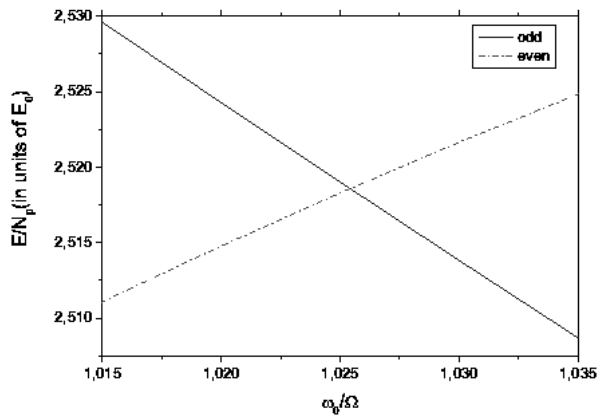}
\caption{Transition from zero to one vortex row for a quasi-parabolic confimente with an infinite potential of width $4a_0$. All the intermediate states dissapear from the transition. \label{fig6}}
\end{figure} 

\begin{figure}
\centering
\includegraphics[scale=0.6]{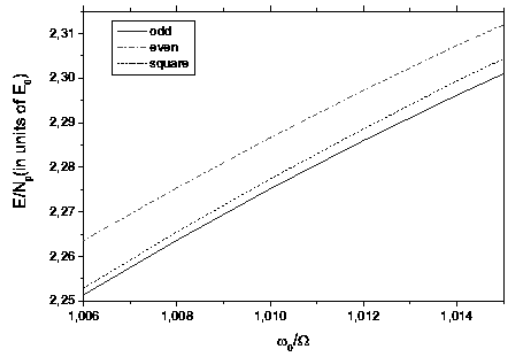}
\caption{The effects of the infinite confinement produces that non triangular configurations appears. However the square lattice is not a ground state of the system.\label{fig5}}
\end{figure}

\begin{figure}
\centering
\includegraphics[scale=0.2]{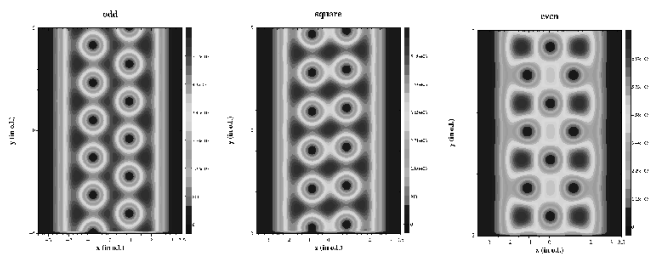}
\caption{Densities for the condensate with an infinite well of width $7a_0$, $gN=1.91$, and $\frac{\omega_0}{\Omega}=1.001$. Lengths are in oscillator units (o.l.). These configurations correspond to those in Fig. \ref{fig5}. \label{densquare}}
\end{figure}

For the well of width $4a_0$ the transition from no vortex to one vortex row losses all the intermediate ground states (see Fig. \ref{fig6}). These results lead us to conclude that the infinite potential suppresses intermediate configurations between transitions that changes the number of rows by one. We also find that the effects of the boundary in the condensate favors configurations without regular triangular form. For example, when the condensate rotation frequency is close to the critical frequency, the extra confinement induces states with a quasi-square lattice of vortices with energies near the triangular ground state. For pure parabolic confinement this state does not appear in our calculations. Although a square lattice was found experimentally in non-equilibrium elongated systems\cite{Engels}, here, the square lattice only appears when an aditional confinement potential is added to the condensate and it does not seem to become the ground state of the system for as far as we have been able to check. The functional energies and the densities of these states are shown in Figs. \ref{fig5} and \ref{densquare}.

\begin{figure}
\centering
\includegraphics[scale=0.3]{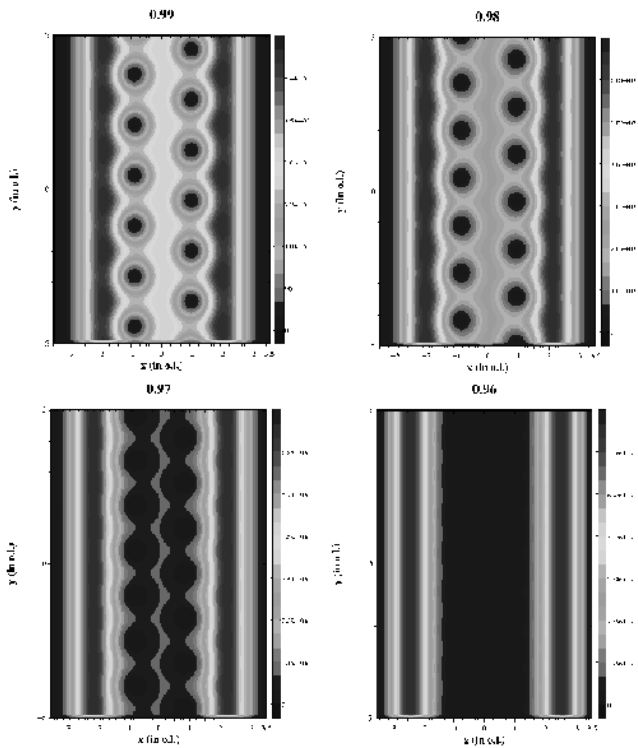}
\caption{BEC densities for $\frac{\omega_0}{\Omega}$=0.99, 0.98, 0.97, and 0.96. Lengths are in oscillator units (o.l.) and $gN=1.91$.\label{densgiga}}
\end{figure}

For rotating frequencies greater than $\omega_0$ the trapping potential has the form of an inverted parabola. However, for frequencies close to the critical frequency the triangular vortex lattice is still the ground state. Finally for ultrafast rotation the BEC splits into two channels at the edges. This is shown in Fig. \ref{densgiga} for a $7a_0$ infinite potential and frequencies $\frac{\omega_0}{\Omega}$=0.99, 0.98, 0.97 and 0.96. As shown, this transition presents strong similarities with the second-order phase transition that takes place in superconducting strips at the onset of surface superconductivity\cite{Palacios:prb:98}.
 
\section{Conclusions}
We have minimized the Gross-Pitaevskii functional with the usual particle number constraint for the problem of a BEC in an infinitely long rotating channel. We find the appearance of vortex rows in the BEC along the longitudinal direction as the rotation increases. The solutions alternate between those with a vortex row in the center of the channel and those with a maximum in the BEC density. For simple parabolic confinement we find intermediate states with non-trivial topologies. When an infinite potential is added at the boundary of the BEC, this intermediate states do not appear in the ground state anymore. This extra confinement produces higher energy solutions in the form of a square vortex lattice and for ultrafast rotation the BEC splits into two channels at the edges.

\section{Acknowledments}
We acknowledge S. Sinha for his suggestions in the initial stages of this work. When preparing this manuscript, we became aware of a very related work by Sinha and Shlyapnikov\cite{Sinha}. The authors acknowledge J. Fern\'andez-Rossier, N. Barber\'an and M. A. Cazalilla for discussions. PS-L thanks the Secretariado de Relaciones con Am\'erica Latina - Centro de Estudios Iberoamericanos Mario Benedetti de la Universidad de Alicante and the Caja de Ahorros del Mediterr\'aneo for its support. This work has been funded by the MCYT under grant MAT2002-04429 and by the Universidad de Alicante.

\section{Appendix I}
The differential equation for the expansion functions $\chi_k$ is
\begin{equation}
-\frac{d^2}{dx^2}\chi_k + \left[ \left(\sqrt{2 R}x-\frac{2}{\sqrt{2R}}\frac{\Omega}{\omega_0}k \right)^2 - \epsilon^*_k \right] \chi_k = 0,
\end{equation}
with $\epsilon^*_k=\epsilon_k-k^2\frac{\left( 1-\frac{\Omega^2}{\omega_0^2} \right)}{\left( 1+\frac{\Omega^2}{\omega_0^2} \right)}$, $\epsilon_k$ the eigenenergies, and $R=1+\frac{\Omega^2}{\omega^2_0}$.
Making a change of variable $u=a\left( \sqrt{2R}x-\frac{2}{\sqrt{2R}}\frac{\Omega}{\omega_0}k\right)$, we rewrite the equation as
\begin{equation}
2Ra^2\frac{d^2}{du^2}\chi_k + \left(\epsilon^*_k-\frac{u^2}{a^2} \right) \chi_k = 0,
\end{equation}
where $a$ is a constant. 
Choosing $a=\sqrt[4]{\frac{1}{2R}}$, the equation reduces to
\begin{equation}
\frac{d^2}{du^2}\chi_k + \left(\frac{\epsilon^*_k}{\sqrt{2R}}-u^2 \right) \chi_k = 0.
\end{equation}
The last equation is similar to the quantum harmonic oscillator equation, so we identify the analytical solutions for the eigenfuctions and the eigenenergies easily:
\begin{equation}
\chi_k = Ce^{-\frac{u^2}{2}}H_n(u)
\end{equation}
and
\begin{equation}
\epsilon^*_k = (2n+1)\sqrt{2R},
\end{equation}
with $H_n$ the Hermite polynomials of grade $n$ and $C$ a normalization constant.
In the lowest Landau level, $(n=0)$, the solutions are finally
\begin{equation}
\chi_k = \sqrt[8]{\frac{2R}{\pi^2}}e^{-\sqrt{\frac{R}{2}}\left(x-\frac{\Omega}{R \omega_0}k \right)^2}
\end{equation}
and
\begin{equation}
\epsilon_k = \sqrt{2R}+k^2 \left(\frac{1-\frac{\Omega^2}{\omega_0^2}}{1+\frac{\Omega^2}{\omega_0^2}} \right).
\end{equation}

As we expected the dispersion relation shows that if the rotating frequency approaches the harmonic frequency $(\Omega \approx \omega_0)$ the parabolic confinement dissapears for all the wavevectors $k$.

\section{Appendix II}
For one component $(N_c=1)$ we have a straightforward minimization process. The functional is
\begin{equation}
E= \epsilon_k |C_k|^2 + \frac{g}{2} |C_k|^4 I_k
\end{equation}
with $I_k=\int \chi_k^4 dx$. Imposing the number of atoms restriction
\begin{equation}
E= (\epsilon_k-\mu) |C_k|^2 + \frac{g}{2} |C_k|^4 I_k +\mu N
\end{equation}
and minimizing with respect to the coefficient $C_k$ we obtain
\begin{equation}
C_k = \sqrt{N},
\end{equation}
\begin{equation}
E_{min} = \epsilon_k N + \frac{g}{2} N^2 I_k,
\end{equation}
and
\begin{equation}
\mu = \epsilon_k + gN I_k,
\end{equation}
with $N=\frac{N_p}{2\pi \frac{a_0}{l}}$ and $N_p$ is the number of particles. For this case the ground state corresponds to $k=0$. We repeat this process for several values of $N_c$ and we select the lowest $E_{min}$ as ground state.

For a two-term linear combination $(N_c=2)$ the constraint is $N=2|C_k|^2$. Using the symmetry we suppose that the state which minimizes the functional is composed by the states $(k,-k)$ and the coefficients $C_k$ and $C_{-k}$ are equal. According to these assumptions
\begin{equation}
E= 2(\epsilon_k - \mu) C_k^2 + g |C_k|^4 A_k + \mu N,
\end{equation}
where $A_k=I_k+2I_{k,-k}$ and $I_{k,-k}=\int dx \chi_k^2 \chi_{-k}^2$. Using the number of particles constraint and minimizing respect the coefficients $C_k$ we find
\begin{equation}
E_{min}= \epsilon_k N + \frac{g}{4} A_k N^2.
\end{equation}
The solutions for the coefficient and the chemical potential are
\begin{equation}
C_k= \sqrt{\frac{N}{2}},
\end{equation}
and
\begin{equation}
\mu= \epsilon_k + \frac{g}{2} A_k N.
\end{equation}
Finding the value of $k$ that minimizes $E_{min}$, we can establish the the density of the BEC. In this case the BEC profile is one centered vortex row.
The last possible analytical calculus is with $N_c=3$. For this configuration we consider a solution $(k,0,-k)$ with the same value for the coefficients $C_k$ and $C_{-k}$ as in $N_c=2$ case, but including a relative phase for the coefficient $C_0=|C_0|e^{i\phi_0}$.
The energy functional with the imposed constraint ($2C_k^2+|C_0|^2=N$) is
\begin{eqnarray}
E & = & 2(\epsilon_k-\mu) C_k^2 + (\epsilon_0-\mu)|C_0|^2 +\frac{g}{2} \nonumber \\
  &   & \left( 2A_kC_k^4 +I_0|C_0|^4 + 4S_{k,\phi_0}C_k^2 |C_0|^2 \right) +\mu N
\end{eqnarray}
with $S_{k,\phi_0}=2I_{k,0}+\cos(2\phi_0)I_{k,0,-k}$ and where a new type of overlap integral between the states, $I_{k,0,-k}=\int dx \chi_k \chi_{-k} \chi_0^2$, appears and represents the correlation between the vortex rows in the condensate. The phase $\phi_0$ which minimizes the functional is $\phi_0=\frac{\pi}{2}$.
Calling $S_k=2I_{k,0}-I_{k,0,-k}$ the three-term functional becomes
\begin{eqnarray}
E & = & 2(\epsilon_k-\mu) C_k^2 + (\epsilon_0-\mu)|C_0|^2 +\frac{g}{2} \nonumber \\
  &   & \left( 2A_kC_k^4 +I(0)|C_0|^4 + 4S_kC_k^2 |C_0|^2 \right) +\mu N.
\end{eqnarray}
Minimizing the functional with respect the coefficients we find
\begin{equation}
C_k^2=\frac{\epsilon_k-\epsilon_0+gN(S_k-I(0))}{g(4S_k-2I(0)-A_k)},
\end{equation}
\begin{equation}
|C_0|^2=\frac{2\epsilon_k-2\epsilon_0+gN(2S_k-A_k)}{g(4S_k-2I(0)-A_k)},
\end{equation}
\begin{equation}
\mu=\frac{2\epsilon_k(S_k-I(0))+\epsilon_0(2S_k-A_k)+gN(2S_k^2-I(0)A_k)}{4S_k-2I(0)-A_k},
\end{equation}
and the minimized functional becomes
\begin{eqnarray}
E_{min}&=&\frac{(\epsilon_k-\epsilon_0)^2}{g(4S_k-2I(0)-A_k)} + \nonumber \\
       & &\frac{N(2\epsilon_k[S_k-I(0)]+\epsilon_0[2S_k-A_k])}{4S_k-2I(0)-A_k}+\nonumber \\
       & &\frac{gN^2(2S_k^2-I(0)A_k)}{2(4S_k-2I(0)-A_k)}.
\end{eqnarray}
As far as the chemical potential is concerned
\begin{equation}
\frac{\partial E_{min}}{\partial N}= \mu,
\end{equation}
is easy to see that satifies the expression
\begin{equation}
\mu=\frac{E_{kin}+E_{ho}+E_{rot}+2E_{int}}{N}
\end{equation}
found after the direct integration of the GP equation, where $E_{kin}$, $E_{ho}$, $E_{rot}$ and $E_{int}$ are the kinetic, harmonic oscillator, rotational and interaction energies. For three components
\begin{equation}
E_{kin}+E_{ho}+E_{rot}=2 \epsilon_k C_k^2 + \epsilon_0 |C_0|^2,
\end{equation}
and the interaction energy is
\begin{equation}
E_{int}=\frac{g}{2}(2A_kC_k^4+I(0)|C_0|^4+4S_kC_k^2|C_0|^2)
\end{equation}
Replacing the analytical coefficients we recover the chemical potential expression found from the minimization.
We try minimize the functional for 4 o more terms semi-analytically but our results only match those obtained numerically when the overlap between terms produces vortex rows. We think that this happens because of the confinement trapping that maintains the states $k$ close to each other. Thus we perform a numerical minimization keeping our restrictions about the behavior of the coefficients and the phases that relate them.

\end{document}